\documentclass[11pt,a4paper]{article}
\usepackage[utf8]{inputenc}
\usepackage{amsmath}
\usepackage{amsfonts}
\usepackage{amssymb}
\usepackage{graphicx}
\usepackage{geometry}
\usepackage[authoryear]{natbib}
\usepackage{url}
\usepackage{booktabs}
\usepackage{longtable}
\usepackage{array}
\usepackage{hyperref}
\usepackage{xurl}             
\hypersetup{hidelinks,breaklinks=true}
\hypersetup{breaklinks=true}
\usepackage[htt]{hyphenat} % allows hyphenation in all-caps words
\title{An Error Model for Evaluating the Accuracy of Satellite-Based XCO$_2$ Products}

\author{Vineet Yadav\textsuperscript{1\#}, Jon Hobbs\textsuperscript{1}, Hai M Nguyen\textsuperscript{1}, Susan S. Kulawik\textsuperscript{2}, \\
Junjie Liu\textsuperscript{1}, David F. Baker\textsuperscript{3}, Isamu Morino\textsuperscript{4}, Hirofumi Ohyama\textsuperscript{4}, \\
Voltaire A. Velazco\textsuperscript{5}, Mihalis Vrekoussis\textsuperscript{6,7,8}, Manvendra K. Dubey\textsuperscript{9}}

\date{}

\begin{document}

\maketitle

\noindent
\textsuperscript{1}Jet Propulsion Laboratory, California Institute of Technology, 4800 Oak Grove Drive, Pasadena, California \\
\textsuperscript{2}Bay Area Research Institute, Moffett Field, California, USA \\
\textsuperscript{3}Cooperative Institute for Research in the Atmosphere, Colorado State University, Fort Collins, Colorado \\
\textsuperscript{4}National Institute for Environmental Studies (NIES), Tsukuba, 305-8506, Ibaraki, Japan \\
\textsuperscript{5}Deutscher Wetterdienst (DWD), Meteorological Observatory Hohenpeissenberg, 82383 Hohenpeissenberg Germany \\
\textsuperscript{6}Climate and Atmosphere Research Center (CARE-C), The Cyprus Institute, Nicosia, Cyprus \\
\textsuperscript{7}Institute of Environmental Physics and Remote Sensing (IUP), University of Bremen, Germany \\
\textsuperscript{8}Center of Marine Environmental Sciences (MARUM), University of Bremen, Bremen (IUP), University of Bremen, Germany \\
\textsuperscript{9}Los Alamos National Laboratory, Los Alamos, New Mexico, United States

\begin{abstract}
Several satellites (e.g., OCO-2 \& 3) and their derived products now provide spatially extensive coverage of the abundance of carbon dioxide in the atmospheric column (XCO$_2$). However, the accuracy of the XCO$_2$ reported in these products needs to be carefully assessed for any downstream scientific analysis; this involves comparison with reference datasets, such as those from the Total Carbon Column Observing Network (TCCON). Previously, systematic and random errors have been used to quantify differences between satellite-based XCO$_2$ measurements and TCCON data. The spatiotemporal density of satellite observations enables the decomposition of the error variability into these components. This study aims to unify the definitions of these error components through a hierarchical statistical model with explicit mathematical terms, which enables a formal definition of the underlying assumptions and estimation of each component. Specifically, we focus on defining model elements, like global bias and systematic and random error, as part of this framework. We use it to compare OCO-2 XCO$_2$ v11.1 data (both original scenes from the `Lite' files and 10-sec averages) and gridded Making Earth System Data Records for Use in Research Environments (MEaSUREs) products to TCCON data. The MEaSUREs products exhibit comparable systematic errors to other OCO-2 products, with larger errors over land versus ocean. We describe the methodology for creating the MEaSUREs products, including their prior and posterior error covariances, with information on spatial correlation for efficient incorporation into scientific analysis.
\end{abstract}

\let\thefootnote\relax\footnotetext{\copyright 2025. All rights reserved.}

\clearpage

\section{Introduction}

Satellite-based measurements of column-averaged dry air mixing ratios of carbon dioxide (XCO$_2$), when combined with \emph{in situ} and aircraft CO$_2$ measurements, can help reduce the sparsity of measurements, uncertainty in the CO$_2$ fluxes estimated from these data through data assimilation (DA), and under-determinedness (degrees of freedom) of the grid-scale estimates of CO$_2$ fluxes derived from these data. Thus, satellite data provide an opportunity to infer finer-scale carbon dioxide (CO$_2$) fluxes than \emph{in situ} CO$_2$ data alone, potentially improving modeling, monitoring, and understanding of CO$_2$ sources and sinks \citep{rayner2001utility}.

While multiple satellites now measure XCO$_2$, their utilization in DA systems or other scientific analyses necessitates a rigorous characterization of the errors associated with these measurements. This characterization is mainly obtained by comparing them with ground-based XCO$_2$ data from the Total Carbon Column Observing Network (TCCON) and CO$_2$ profile measurements taken from aircraft flights. The development and continued upkeep of the TCCON network provide a regular source of validation data for satellite-based estimates of atmospheric greenhouse gases. The network's observations are benchmarked against coincident aircraft campaigns \citep{wunch2010}, providing a critical link in a validation strategy to connect satellite products to established global CO$_2$ standards. TCCON has been a consistent source for multi-year validation data for the Greenhouse Gases Observing Satellite (GOSAT; \citealt{Wunch2011PTRSA}) and the Orbiting Carbon Observatory-2 (OCO-2; \citealt{wunch2017}) mission. This extended record enables a common framework for validation to be applied across multiple data products from multiple instruments \citep{kulawik2016}.

This validation framework includes a methodology for partitioning the satellite retrieval errors into different components that span different spatial and temporal scales. This partitioning is particularly informative for satellite products with fine spatial resolution and global coverage, with retrieval errors that vary in space and time. This is an important consideration for OCO-2, which provides a multi-year record of XCO$_2$ estimates with global coverage. Validation of these products has revealed a combination of single-retrieval random errors with coherent errors at local and regional spatial scales \citep{kulawik2019a, worden2017}. This combination of errors has implications for the statistical properties of estimates that combine multiple soundings together and for the use of products from multiple satellites \citep{taylor2023}.

For conceptual and implementation clarity, the validation formalism developed by \citet{kulawik2016} would benefit from an underlying probabilistic theoretical basis and consistent notation. To address this, we have developed a hierarchical statistical framework that can be replicated across studies, and we apply it here to compute component errors associated with three products based on satellite XCO$_2$ measurements.

This statistical framework is essential because combining XCO$_2$ measurements from multiple satellites to create merged products increases measurement density and facilitates scientific analysis. However, proper statistical assessment of errors of these products is necessary since the original products used in creating the merged product have different observational geometries, densities, footprints, and sometimes different averaging kernels and pressure weighting functions. As a result, the atmospheric column corresponding to the XCO$_2$ reported in these merged or derived products remains indeterminate. This indeterminacy arises because these merged products are generated by combining measurements taken at different but proximate spatiotemporal locations and from instruments with varying sensitivities to different portions of the atmospheric column. Therefore, it is crucial to carefully manage the associated uncertainties to ensure a robust pipeline for scientific analysis of the spatiotemporal patterns of XCO$_2$.

Examples of derived/merged products are the MEaSUREs (Making Earth System Data Records for Use in Research Environments) XCO$_2$ products that use Gaussian processes and Vecchia approximation (e.g., \citealt{Katzfuss2020}) to generate fused, gridded estimates of XCO$_2$ \citep{nguyen2012} while accounting for variations in sensor geometries, revisit times, and measurement accuracies of instruments. These products are unique in the sense that they not only come with a gridded estimate of XCO$_2$ at a different spatial resolution than the base products but also include at that resolution the pressure weighting function, averaging kernel, and Coordinated Universal Time of observations and, for the first time, include a full posterior error covariance.

However, no matter how advanced the statistical methods used to create these gridded products are, they still need to be validated. This validation is crucial to ensure the reliability and accuracy of the derived and merged products.

Given this background, this study has two objectives. The first is to formalize and document a hierarchical statistical model to underlie the methodology used previously for validating satellite-based XCO$_2$ products (see \citealt{kulawik2016}), where the reference XCO$_2$ data are those available from ground-based sensors. The second objective is to formulate a spatial statistical model for fusing XCO$_2$ products from multiple sensors that produce full prior and posterior error covariance.

The first objective does not require that the reference dataset be based only on ground-based sensors; instead, it is a general framework that can be applied to compare XCO$_2$ products with any reference dataset. The second objective is linked to the first objective insofar as it outlines the methodology for creating merged or gridded products with full posterior error covariance that can be included in downstream scientific analyses and tests the product's performance within the error assessment framework developed as part of the first objective.

This paper is organized as follows. In Section~\ref{sec:data}, we describe the satellite-derived XCO$_2$ products that are evaluated. These products include the MEaSUREs fused XCO$_2$ estimates. Section~\ref{sec:methodology} describes the validation methodology and the associated hierarchical statistical model. The development distinguishes between systematic and random errors and the implications for aggregated XCO$_2$ estimates, such as the fused products. The section also describes the spatial statistical data fusion methodology used to produce the MEaSUREs products. The validation results are reported in Section~\ref{sec:results}, and Section~\ref{sec:conclusion} provides concluding remarks.

\section{Data: Satellite XCO$_2$ Products used for comparison with TCCON reference dataset}
\label{sec:data}

All data used in this work are accessible online. Using the error framework outlined in the methods section, we compare four satellite-based data products to the ground-based TCCON data. TCCON is a network of ground-based Fourier Transform Spectrometers recording direct solar spectra in the near-infrared spectral region located around the globe. These TCCON sites provide accurate and precise column-averaged abundances of CO$_2$ among measurements of other trace gas species (see \citealt{wunch2025}). TCCON data are available from \url{https://tccondata.org/2020} \citep{TCCONTeam2022}, and the data or manuscript references associated with data for each site are given in supplementary material.

While the methodology for creating two satellite-based XCO$_2$ products has been detailed in previous publications, the methods behind MEaSUREs products have not been described; therefore, we outline the process for creating these in Section~\ref{sec:methodology}.

The first satellite-based product we used for comparison is the OCO-2 single-sounding bias-corrected XCO$_2$, provided through the OCO-2 mission's ``Lite'' data product, LtXCO$_2$ v11.1 (DOI: 10.5067/8E4VLCK16O6Q, \citealt{OCO2Team2022}), a standard OCO-2 product archived at the NASA Goddard Earth Sciences Data and Information Services Center (GESDISC). The mission's data user guide describes the variables provided in the products \citep{payne2023}, and \citet{taylor2023} describe the product's spatiotemporal coverage and the retrieval algorithm, known as the Atmospheric Carbon Observations from Space (ACOS) algorithm.

The second product, OCO-2 10-sec \citep{baker2024}, is described by \citet{peiro2022}. This product provides XCO$_2$ estimates aggregated at 10-second intervals based on the OCO-2 single-sounding retrievals. It also includes estimates of the uncertainty in the 10-second averages computed to account for correlations in the errors of the averaged values \citep{baker2022}. This aggregation yields a reduced data volume suitable for input into global flux inversion systems.

Two gridded products were created as part of MEaSUREs work. These products utilize OCO-2 LtXCO$_2$ v11.1 (for generic details about OCO-2 products, see \citealt{Crisp2020, odell2012}) and ACOS-GOSAT XCO$_2$ v9.0 data (see \citealt{morino2011}). We refer to these products as (1) MEaSUREs OCO-2 (DOI: 10.5067/582L7HTJ343N, \citealt{Nguyen2022b_dataset}) and (2) MEaSUREs OCO-2 and GOSAT (DOI: 10.5067/ZS346LH1NTIS, \citealt{Nguyen2022a_dataset}).

The MEaSUREs OCO-2 product is a gap-filled product obtained from fusing Level 2 bias-corrected XCO$_2$ retrievals from OCO-2 LtXCO$_2$ v11.1, whereas the MEaSUREs OCO-2 and GOSAT product combines OCO-2 LtXCO$_2$ v11.1 and ACOS-GOSAT XCO$_2$ v9.0 to create a fused product.

These two MEaSUREs products have global coverage at 1° spatial resolution and a daily temporal cadence. The MEaSUREs OCO-2 and GOSAT fused product has temporal coverage from September 2014 to December 2019, the last month the ACOS-GOSAT v9.0 product is available. The OCO-2-based gridded product is being processed continuously and is available from September 2014 onwards.

\section{Methodology}
\label{sec:methodology}

In the coming sections, we will distinguish TCCON XCO$_2$ data from satellite-based XCO$_2$ data. We always use the word or acronym `XCO$_2$' to indicate satellite-based observations, while all TCCON-based XCO$_2$ data will just be called `TCCON data' unless specifically mentioned otherwise. We first describe the methodology for creating MEaSUREs products. This is followed by the hierarchical statistical error model to assess all four products mentioned in the previous section.

\subsection{MEaSUREs Product: Development Methodology}

We use an overarching framework of Gaussian processes to fuse XCO$_2$ observations from multiple instruments to create MEaSUREs products. Specifically, we combine Kriging with Vecchia's approximation of Gaussian processes to create a gridded and fused dataset. Mathematically, the methodology to combine multiple datasets applies similarly to one or multiple datasets \citep{nguyen2012}. To simplify the notation, we describe the mathematical framework to combine observations from multiple instruments in terms of observations available from a single instrument.

\subsubsection{Kriging Framework}

Given a vector $\mathbf{z}$ of XCO$_2$ observations such that:
\begin{equation}
\label{eq:kriging1}
\mathbf{z} = \left(z(\mathbf{s}_1), z(\mathbf{s}_2), z(\mathbf{s}_3), \ldots, z(\mathbf{s}_p)\right)
\end{equation}

where $z(\mathbf{s}_i)$ represents an individual XCO$_2$ observation at a location $\mathbf{s}_i$ in space and time (i.e., a vector consisting of latitude, longitude, and time). The $\mathbf{z}$ data vector can consist of XCO$_2$ observations from one or many instruments. Further, an $i$-th XCO$_2$ observation can be partitioned as:
\begin{equation}
\label{eq:kriging2}
z(\mathbf{s}_i) = Y(\mathbf{s}_i) + \epsilon(\mathbf{s}_i)
\end{equation}

where $z(\mathbf{s}_i)$ is the sum of true unobserved XCO$_2$ $Y(\mathbf{s}_i)$, plus error $\epsilon(\mathbf{s}_i)$. Note that if data vector $\mathbf{z}$ consists of XCO$_2$ observations from two or more instruments, then $\epsilon(\mathbf{s}_i)$ may vary based on instruments.

We can obtain an estimate of XCO$_2$ i.e., $\hat{Y}$ at an unobserved location $\mathbf{s}_0$ as:
\begin{equation}
\label{eq:kriging3}
\hat{Y}(\mathbf{s}_0) = \mathbf{a}_0^T \mathbf{z}
\end{equation}

where $\mathbf{a}_0$ is a $p$-dimensional vector of kriging coefficients associated with location $\mathbf{s}_0$. The vector $\mathbf{a}_0$ in Eq.~\eqref{eq:kriging3} can be obtained by minimizing:
\begin{align}
\label{eq:kriging4}
E\left[(Y(\mathbf{s}_0) - \hat{Y}(\mathbf{s}_0))^2\right] &= \text{Var}(Y(\mathbf{s}_0) - \mathbf{a}_0^T\mathbf{z})\\
E\left[(Y(\mathbf{s}_0) - \hat{Y}(\mathbf{s}_0))^2\right] &= \text{Var}(Y(\mathbf{s}_0)) - 2\mathbf{a}_0^T\text{Cov}(\mathbf{z}, Y(\mathbf{s}_0)) + \mathbf{a}_0^T\text{Var}(\mathbf{z})\mathbf{a}_0\nonumber
\end{align}

with respect to $\mathbf{a}_0$, subject to the unbiasedness constraint i.e.,
\begin{equation}
\label{eq:kriging5}
1 = \mathbf{a}_0^T \mathbf{1}
\end{equation}

Note that this vector of kriging coefficients in vector $\mathbf{a}_0$ is precisely required for forming linear combinations of fields such as averaging kernels and XCO$_2$ priors. We can solve the minimization problem in Eq.~\eqref{eq:kriging4} to obtain the optimal $\mathbf{a}_0$ using the Lagrange multiplier method, and the resulting solutions are known as the Kriging Equations (for details, see \citealt{cressie2015}).

\subsubsection{Vecchia approximation of Kriging Covariances}

The solution of the minimization in Eq.~\eqref{eq:kriging4} requires inverting of a $(p + 1) \times (p + 1)$ matrix $\text{Cov}(\mathbf{z})$, where $p$ is the length of the data vector $\mathbf{z}$. This inversion has a computational complexity $O((p + 1)^3)$ hence a brute force solution of Eq.~\eqref{eq:kriging4} is not possible for large $p$. Therefore, we use a framework outlined in \citet{Katzfuss2020}, which approximates the inverse of $\text{Cov}(\mathbf{z})$ required for obtaining a solution of Eq.~\eqref{eq:kriging4} using a technique called Vecchia approximation.

To reiterate, a significant bottleneck in the formulation above is the inversion of the joint covariance matrix $\boldsymbol{\Sigma} = \text{Cov}(\mathbf{z})$, where the size of $\boldsymbol{\Sigma}$ can be on the order of tens of thousands. The key to the Vecchia approximation is the joint probability distribution of the data, denoted $P(\mathbf{z})$, can be described with through a series of conditional probabilities.

As an illustration, consider a simplified problem where $\mathbf{z} = (z(\mathbf{s}_1), z(\mathbf{s}_2), z(\mathbf{s}_3))$, then 
\begin{equation}
\label{eq:vecchia1}
P(\mathbf{z}) = P(z(\mathbf{s}_1), z(\mathbf{s}_2), z(\mathbf{s}_3))
\end{equation}

which is equivalent to:
\begin{equation}
\label{eq:vecchia2}
P(\mathbf{z}) = P(z(\mathbf{s}_1))P(z(\mathbf{s}_2)|z(\mathbf{s}_1))P(z(\mathbf{s}_3)|z(\mathbf{s}_1), z(\mathbf{s}_2))
\end{equation}

where all terms are as defined earlier.

The Vecchia approximation of Eq.~\eqref{eq:vecchia2} takes the form of:
\begin{equation}
\label{eq:vecchia3}
P(\mathbf{z}) \approx P(z(\mathbf{s}_1))P(z(\mathbf{s}_2)|z(\mathbf{s}_1))P(z(\mathbf{s}_3)|z(\mathbf{s}_2))
\end{equation}

where we assume that each location is conditionally independent of the entire dataset given knowledge of some locations in the local spatiotemporal neighborhood. \citet{vecchia1988} showed that this approximation allows for the computation of the sparse Cholesky factor of the precision matrix. The Vecchia approximation enables us to directly compute the inverse of $\text{Cov}(\mathbf{z})$ i.e., $\boldsymbol{\Sigma}^{-1}$, which is required to solve for the kriging vector $\mathbf{a}_0$ in Eq.~\eqref{eq:kriging4}. For more detail on Vecchia approximations of Gaussian-process predictions, see \citet{Katzfuss2020}.

In this work, we use the GPVecchia package available from (\url{https://cran.r-project.org/web/packages/GPvecchia/}) to compute $\boldsymbol{\Sigma}^{-1}$ (for details on methodology see, \citealt{Katzfuss2020}). Traditionally, uncertainty estimates of XCO$_2$, such as those provided with OCO-2 and ACOS, only focus on the diagonal of the full covariance matrix and ignore the off-diagonal dependences. These uncertainties are calculated as the diagonal of the posterior covariance matrix, $\boldsymbol{\Sigma}_p = \text{Cov}(\mathbf{Y}_p | \mathbf{z})$, where $\mathbf{Y}_p$ is the set of gridded fused XCO$_2$ outputs. One particularly useful feature of GPVecchia package is that it provides the full sparse posterior precision matrix for the set of fusion locations.

\subsubsection{Fusion of multiple data products}

Thus far we have discussed the case of interpolation of data from a single instrument, which is used to create the MEaSUREs OCO-2 product \citep{Nguyen2022b_dataset}. Here we will give a short overview of its extension to the case of multiple instruments, as applied to the creation of the MEaSUREs OCO-2 and GOSAT product \citep{Nguyen2022a_dataset}. For ease of notation, we assume that we have two instruments, denoted with the corresponding subscript as in the data model below:
\begin{align}
\label{eq:fusion1}
\mathbf{z}_1(\mathbf{s}_{1,i}) &= Y(\mathbf{s}_{1,i}) + \epsilon_1(\mathbf{s}_{1,i})\\
\mathbf{z}_2(\mathbf{s}_{2,i}) &= Y(\mathbf{s}_{2,i}) + \epsilon_2(\mathbf{s}_{2,i})\nonumber
\end{align}

where $\mathbf{z}_1$ is the data vector of XCO$_2$ values from the first instrument (say, OCO-2) and $\mathbf{z}_2$ denote data from the second instrument (e.g., GOSAT). Our goal is to obtain a fused value at a new location as a linear combination of the two data vectors $\mathbf{z}_1$ and $\mathbf{z}_2$ as follows:
\begin{equation}
\label{eq:fusion2}
\hat{Y}(\mathbf{s}_0) = \mathbf{a}_1^T\mathbf{z}_1 + \mathbf{a}_2^T\mathbf{z}_2
\end{equation}

where the coefficients $\mathbf{a}_1$ and $\mathbf{a}_2$ are chosen through an optimization procedure to minimize the expected squared error.
\begin{equation}
\label{eq:fusion3}
E\left[(Y(\mathbf{s}_0) - \hat{Y}(\mathbf{s}_0))^2\right] = \text{Var}(Y(\mathbf{s}_0) - \mathbf{a}_1^T\mathbf{z}_1 - \mathbf{a}_2^T\mathbf{z}_2)
\end{equation}

The key insight in this fusion approach is that we can concatenate $\mathbf{a}_1^T$ and $\mathbf{a}_2^T$ into a new `fusion' vector $\mathbf{a}_F^T$, and similar apply the same concatenation process to $\mathbf{z}_1$ and $\mathbf{z}_1$ to produce $\mathbf{z}_F$. That is, we conceptually combine data from multiple instruments as a single meta-dataset and then apply formalism as the single-instrument case. Under this approach, the equation for the fused value is now written as:
\begin{equation}
\label{eq:fusion4}
\hat{Y}(\mathbf{s}_0) = \begin{bmatrix} \mathbf{a}_1^T, \mathbf{a}_2^T \end{bmatrix} \begin{bmatrix} \mathbf{z}_1 \\ \mathbf{z}_1 \end{bmatrix} = \mathbf{a}_F^T \mathbf{z}_F
\end{equation}

And similarly, the expected squared error to be minimized over can now be written as:
\begin{equation}
\label{eq:fusion5}
E\left[(Y(\mathbf{s}_0) - \hat{Y}(\mathbf{s}_0))^2\right] = \text{Var}(Y(\mathbf{s}_0) - \mathbf{a}_1^T\mathbf{z}_1 - \mathbf{a}_2^T\mathbf{z}_2)
\end{equation}

which has the same format as Eq.~\eqref{eq:kriging3} and \eqref{eq:kriging4}, and thus the same formalism and optimization procedure for the single-instrument case in Section 3.2.1 and 3.2.2 can be applied to the fusion case with multiple input instruments. For more detail on treating multi-instrument fusion as a generalization of single-instrument interpolation, see \citet{nguyen2012, Nguyen2014Technometrics}.

\subsubsection{Covariance and the Precision Information in the Product}

Our product stores the precision matrix $\boldsymbol{\Sigma}_p^{-1}$ in NetCDF files in Dictionary of Keys (DOK, for details, see \url{https://docs.scipy.org/doc/scipy/reference/generated/scipy.sparse.dok_matrix.html}) sparse matrix format, which reduces the file size and allows for rapid reconstruction of the matrix in dense format. We do not store the full covariance matrix $\boldsymbol{\Sigma}_p$, but it is straightforward for the data user to read in $\boldsymbol{\Sigma}_p^{-1}$ and compute its inverse to obtain $\boldsymbol{\Sigma}_p$. For convenience, the uncertainty of $\hat{Y}(\mathbf{s}_0)$ is reported in the NetCDF files as $\text{diag}(\boldsymbol{\Sigma}_p)$.

Under the assumption of normality, either $\boldsymbol{\Sigma}_p$ or $\boldsymbol{\Sigma}_p^{-1}$ can be used to generate multiple realizations of $\mathbf{Y}_p$, which by using cholesky ($chol$) decomposition of $\boldsymbol{\Sigma}_p$ can be given as:
\begin{equation}
\label{eq:cholesky}
\mathbf{Z} = \hat{\mathbf{Y}}_p + chol(\boldsymbol{\Sigma}_p)\mathbf{M}
\end{equation}

where $\mathbf{M}$ is a vector of dimension $p$ that consists of independent and identically distributed (iid) normal random vectors with mean 0 and variance 1 (i.e., N(0,1)) and $\mathbf{Z}$ are realizations of $\mathbf{Y}_p$.

\subsubsection{Workflow to creating fused datasets from multiple instruments}

A sequential procedure was adopted to create gridded and fused datasets. The process begins by dividing the daily global data into a 1° × 1° output grid. Afterward, the OCO-2 and ACOS XCO$_2$ observations are filtered using the provided quality filters and subsetted by Land or Ocean. The filtered data is then processed through the GPVecchia package, and the optimal fusion coefficient vector and the posterior precision matrices are computed separately for Land and Ocean. Additionally, linear combinations of other fields, such as averaging kernels, prior mean profiles, and pressure levels, are calculated using the coefficient vector $\mathbf{a}_0$, ensuring that its elements sum to 1. Finally, the results are stored in a single daily output NetCDF file, with variable details accessible through any NetCDF file viewer. Further information regarding the data generation process and output formats can be found in the Algorithm Theoretical Basis Document and Data User Guide \citep{Nguyen2022UserGuide, Nguyen2022ATBD}.

\subsection{Statistical Error model for validation of satellite based XCO$_2$ observations}

A comparison of XCO$_2$ observations from satellite products to TCCON XCO$_2$ data or other reference datasets requires a statistical model that not only formalizes the partitioning of total error between its constituent components (such as systematic, random, and co-location errors) but also provides an intuitive understanding of these errors.

A hierarchical or multi-level statistical model for the XCO$_2$ observations and a corresponding model for the TCCON measurements can motivate the characterization of the retrieved XCO$_2$ error distributions. To our knowledge, this is the first formalized error model of the OCO-2 satellite observations based on comparing retrieved XCO$_2$ with TCCON measurements.

We define $\hat{x}_{ijk}$ to be a retrieved XCO$_2$ observation from a product (e.g. OCO-2 LtXCO$_2$, OCO-2 10-sec, or MEaSUREs). There are typically multiple soundings/estimates $\hat{x}_{ijk}$; $i = 1, \ldots, n_{jk}$ for a coincidence with TCCON station $j$ on day $k$. We assume there is a single validation vertical profile $\mathbf{x}_{val,jk}$ for a day/station coincidence. For error assessment, the validation profile is convolved with the individual satellite product averaging kernels,
\begin{equation}
\label{eq:validation1}
x_{est,ijk} = \mathbf{h}_{ijk}^T\mathbf{x}_{a,ijk} + \mathbf{h}_{ijk}^T\mathbf{A}_{ijk}(\mathbf{x}_{val,jk} - \mathbf{x}_{a,ijk})
\end{equation}

where $\mathbf{x}_{a,ijk}$ is the a priori CO$_2$ profile, $\mathbf{h}_{ijk}$ is the pressure weighting function, and $\mathbf{A}_{ijk}$ is the averaging kernel matrix. The convolved quantity $x_{est,ijk}$ can be interpreted as how the satellite (or derived product) would have observed the true state given the retrieval/instrument sensitivity.

We adopt a hierarchical, or multi-level, statistical model for describing the distribution of retrieval errors. This model can be written as:
\begin{equation}
\label{eq:hierarchical}
\hat{x}_{ijk} = x_{ijk} + \mu + \alpha_j + \gamma_{jk} + \epsilon_{ijk}
\end{equation}

where, $x_{ijk}$ is the true XCO$_2$ corresponding to the individual retrieved product. The error is then decomposed into an overall product bias $\mu$, a TCCON station-specific bias $\alpha_j$, a daily overpass error $\gamma_{jk}$, and a single retrieval error $\epsilon_{ijk}$. The overall bias is fixed, and the remaining components are random variables that are assumed to be independent with component-specific variances given as:
\begin{align}
\label{eq:variance_components}
\alpha_j &\sim N(0, \sigma_\alpha^2)\\
\gamma_{jk} &\sim N(0, \sigma_\gamma^2)\nonumber\\
\epsilon_{ijk} &\sim N(0, \sigma_\epsilon^2)\nonumber
\end{align}

Since TCCON measurements are themselves uncertain, a complementary model accounting for their uncertainty can be given as:
\begin{align}
\label{eq:tccon_uncertainty}
\mathbf{x}_{val,jk} &= \mathbf{x}_{jk} + \boldsymbol{\epsilon}_{val,jk}\\
x_{est,ijk} &= x_{jk} + \epsilon_{val,jk}^* + \kappa_{ijk}\nonumber
\end{align}

where, $\mathbf{x}_{val,jk}$ is the TCCON CO$_2$ profile, $x_{est,ijk}$ is the TCCON XCO$_2$ value, $\epsilon_{val,jk}^*$ is a measurement error for the TCCON measurement, and $\kappa_{ijk}$ is a local aggregation error incurred from smoothing the TCCON profile with slightly variable averaging kernels within the coincidence region (this error is generally small).

Overall, the error assessment is based on statistical summaries of $\hat{x}_{ijk} - x_{est,ijk}$, which can be broken down conceptually into components,
\begin{alignat}{3}
\label{eq:error_decomposition}
\hat{x}_{ijk} - x_{est,ijk} &= \mu && \quad \text{global bias}\\
&+ x_{ijk} - x_{jk} && \quad \text{co-location}\nonumber\\
&+ \epsilon_{val,jk}^* + \kappa_{ijk} && \quad \text{validation}\nonumber\\
&+ \alpha_j + \gamma_{jk} && \quad \text{systematic}\nonumber\\
&+ \epsilon_{ijk} && \quad \text{random}\nonumber
\end{alignat}

This partitioning in Eq.~\eqref{eq:error_decomposition} incorporates a combination of the retrieval errors from Eq.~\eqref{eq:hierarchical}, separating the overall global bias $\mu$, systematic error components that are constant within an overpass, and random error that is unique to each retrieval. Other error sources, namely validation and co-location, contribute to the difference between the retrieval and the TCCON reference in Eq.~\eqref{eq:error_decomposition}.

Based on partitioning of errors in Eq.~\eqref{eq:error_decomposition} a daily average error $\bar{e}_{jk}$ for station $j$ on day $k$ can be computed as:
\begin{equation}
\label{eq:daily_average}
\bar{e}_{jk} = \frac{1}{n_{jk}}\sum_{i=1}^{n_{jk}} (\hat{x}_{ijk} - x_{est,ijk})
\end{equation}

Application of this averaging to the statistical model in Eq.~\eqref{eq:error_decomposition}, reduces the impact of the individual random errors with increasing sample size $n_{jk}$, while systematic errors remain a sizable contribution to the daily average error as given below:
\begin{alignat}{3}
\label{eq:daily_average_decomposition}
\bar{e}_{jk} &= \mu && \quad \text{global bias}\\
&+ \frac{1}{n_{jk}}\sum_{i=1}^{n_{jk}} x_{ijk} - x_{jk} && \quad \text{co-location}\nonumber\\
&+ \epsilon_{val,jk}^* + \frac{1}{n_{jk}}\sum_{i=1}^{n_{jk}} \kappa_{ijk} && \quad \text{validation}\nonumber\\
&+ \alpha_j + \gamma_{jk} && \quad \text{systematic}\nonumber\\
&+ \frac{1}{n_{jk}}\sum_{i=1}^{n_{jk}} \epsilon_{ijk} && \quad \text{random}\nonumber
\end{alignat}

Overall, the error assessment seeks to isolate $\text{Var}(\alpha_j + \gamma_{jk}) = \sigma_\alpha^2 + \sigma_\gamma^2$, i.e., the variability in systematic error contributions. Conceptually, this is the variability of errors of small area averages. This is not directly available but can be estimated by computing the variances of other components in this decomposition and the overall variance of the daily average errors. In the decomposition shown in Eq.~\eqref{eq:daily_average_decomposition}, it is assumed that all the terms are independent.

Multiple sources of uncertainty contribute to the variances of the systematic and random error components of the error decomposition. Instrument measurement errors and retrieval error cross-correlations, termed interference errors, contribute substantially to the variability in XCO$_2$ error for individual measurements \citep{hobbs2017, kulawik2019a, kulawik2019b}. In addition, retrieval forward model misspecifications contribute to errors that are often coherent for local areas within a single coincidence but vary at larger spatiotemporal scales \citep{connor2016, hobbs2020}.

\subsection{Application of the Statistical Error model for validation of XCO$_2$ data products}

A theoretical statistical framework for comparing XCO$_2$ data products with reference datasets, using TCCON as a reference, was presented in the previous section. However, applying this framework to quantify errors requires establishing criteria for co-locating retrieved XCO$_2$ with true XCO$_2$ from the reference dataset. This is necessary due to differences in the spatiotemporal resolution and footprint between the two datasets.

The quantification of errors in Eq.~\eqref{eq:error_decomposition} requires establishing co-location criteria for comparing retrieved satellite XCO$_2$ observations against TCCON data, which serve as our reference dataset. Co-location of satellite-based XCO$_2$ observations with TCCON typically involves identifying and averaging satellite observations within specific latitudinal, longitudinal, and temporal windows. In this work, these windows were determined by the spatiotemporal density of XCO$_2$ observations and the similarity between atmospheric CO$_2$ sampled by TCCON and XCO$_2$ observations. We consider XCO$_2$ observations coincidental when they fall within a three-by-five-degree latitude-longitude grid cell containing the TCCON site. For temporal matching, we use 90-minute averages of TCCON measurements taken at 15-minute intervals and select the closest match to OCO-2, requiring the mean TCCON measurement to be within 1 hour of the OCO-2 observations. These spatiotemporal criteria were selected by applying the smallest space-time constraints that would still yield sufficient coincident observations.

It's important to note that XCO$_2$ observations meeting these coincidence criteria vary in both space and time, with each retrieved XCO$_2$ observation having unique latitude, longitude, and time coordinates. In contrast, averaged TCCON data have fixed latitude and longitude coordinates but vary in time.

\subsubsection{Validation: Observation based error analysis}

There are $J = 20$ total TCCON stations for the validation over land and fewer stations over the ocean. Note that matchups are available for some subset of the $j = 1, \ldots, J$ stations on any given day $k$; i.e. for OCO-2 overpasses, only data from some TCCON stations are available on a given day. We construct $\mathbf{d}_j$, a vector containing the indices for days $k$ with matchups for station $j$, and $N_j$, the number of days with matchups for station $j$ (i.e. $N_j$ is the number of elements of $\mathbf{d}_j$).

\begin{figure}[h!]
\centering
\includegraphics[width=0.8\textwidth]{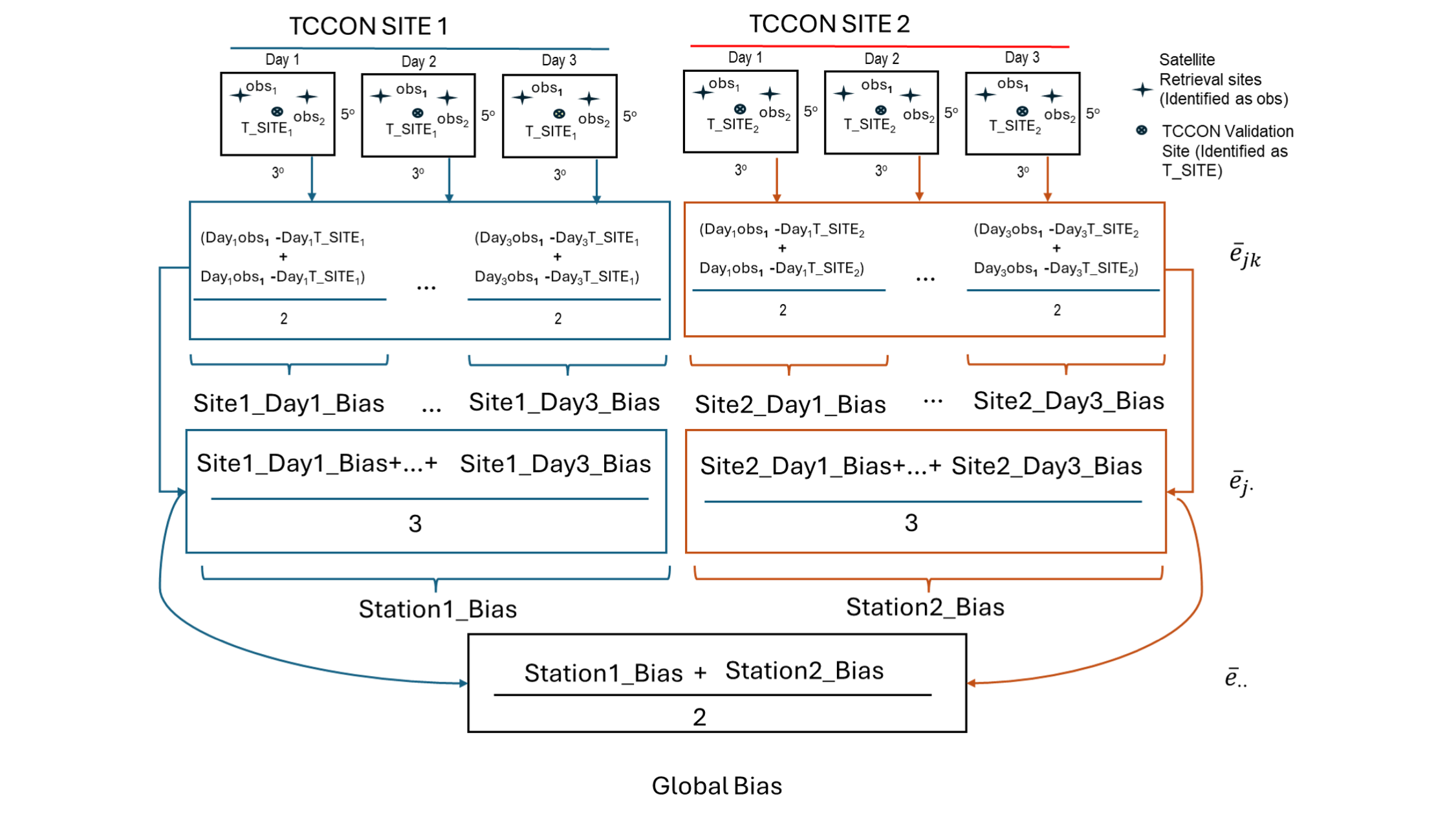}
\caption{A schematic for computing biases at different hierarchies of assessment. The schematic above shows how Global and Station Bias are computed. For demonstration, we assume that two satellite retrievals are available for three days for two TCCON sites. In the results tables, we report Overall Bias or Global Bias and Station Bias. Note that symbols $\bar{e}_{jk}$, $\bar{e}_{j\cdot}$ and $\bar{e}_{\cdot\cdot}$ correspond with Eq.~\eqref{eq:daily_average}, \eqref{eq:station_bias}, and \eqref{eq:global_bias} in the text.}
\label{fig:bias_schematic}
\end{figure}

Given this information, we compute station bias $\bar{e}_{j\cdot}$ (see Figure~\ref{fig:bias_schematic}) for station $j$, which is the mean of daily average errors for that station, which can be written as:
\begin{equation}
\label{eq:station_bias}
\bar{e}_{j\cdot} = \frac{1}{N_j} \sum_{k \in \mathbf{d}_j} \bar{e}_{jk}
\end{equation}

Once station bias has been computed, an overall (global) bias $\bar{e}_{\cdot\cdot}$, which is an average of station biases, is computed (see Figure~\ref{fig:bias_schematic}). The overall bias is defined as:
\begin{equation}
\label{eq:global_bias}
\bar{e}_{\cdot\cdot} = \frac{1}{J} \sum_{j=1}^J \bar{e}_{j\cdot}
\end{equation}

\begin{figure}[h!]
\centering
\includegraphics[width=0.8\textwidth]{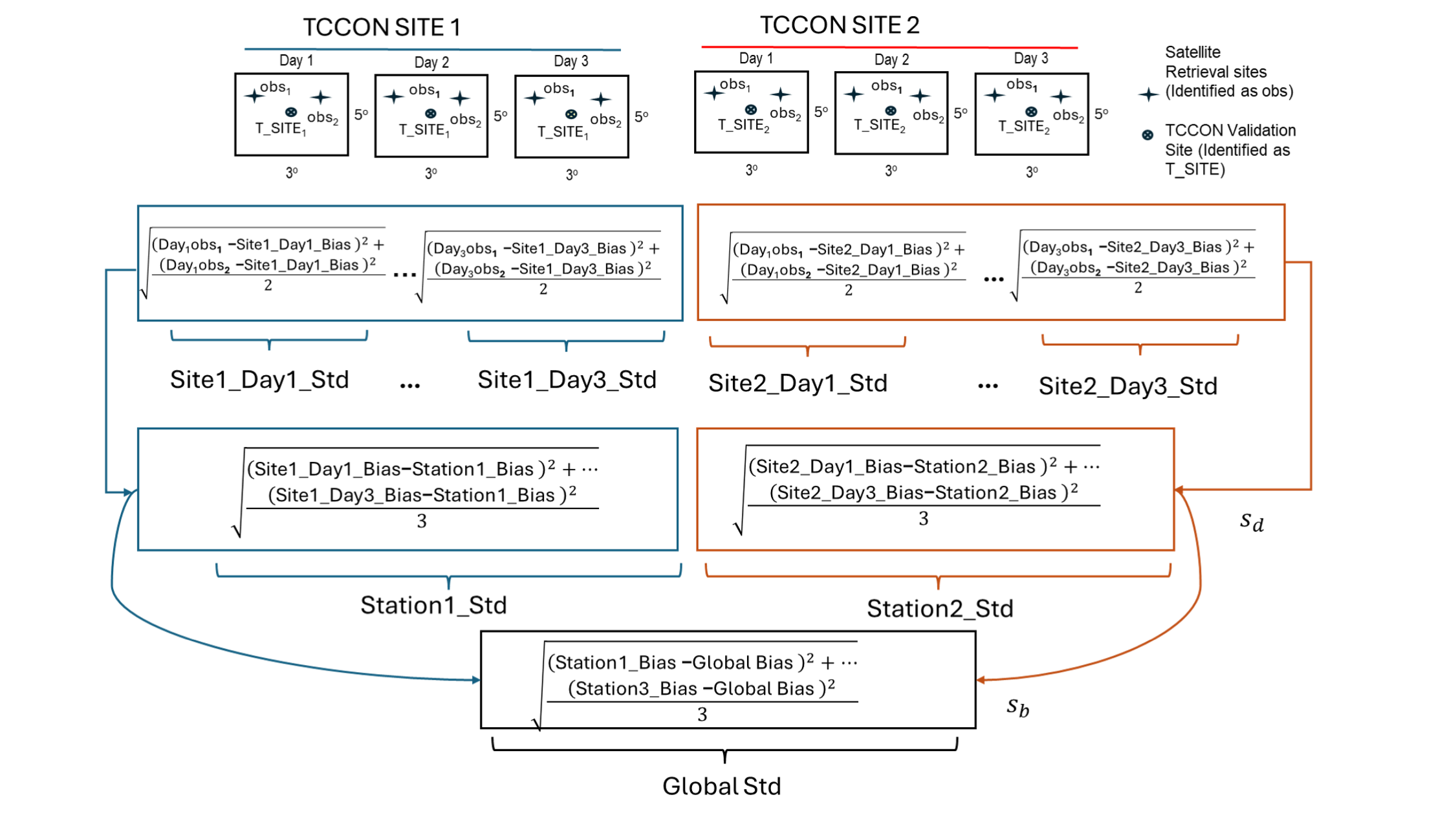}
\caption{A schematic for computing standard deviation (Std) at different assessment hierarchies from biases shown in Figure~\ref{fig:bias_schematic}. These Std's are used in computing systematic errors. Like in Figure~\ref{fig:bias_schematic} for demonstration, we assume that two satellite retrievals are available for three days for two TCCON sites. In the results tables, we report systematic errors that require quantities $s_d$ and $s_b$, which are described as Eq.~\eqref{eq:daily_std} and \eqref{eq:station_std} in the text.}
\label{fig:std_schematic}
\end{figure}

Note that $\bar{e}_{j\cdot}$ is an estimator of the station bias ($\mu + \alpha_j$), $\bar{e}_{\cdot\cdot}$ is an estimator of the overall product bias $\mu$. After computation of biases, Daily Bias Std ($s_d$), and Station Bias Std ($s_b$) are computed (see Figure~\ref{fig:std_schematic}) and these can be given as:
\begin{equation}
\label{eq:daily_std}
s_d = \frac{1}{J}\sum_{j=1}^J \sqrt{\frac{1}{N_j - 1} \sum_{k \in \mathbf{d}_j} (\bar{e}_{jk} - \bar{e}_{j\cdot})^2}
\end{equation}

\begin{equation}
\label{eq:station_std}
s_b = \sqrt{\frac{\sum_{j=1}^J (\bar{e}_{j\cdot} - \bar{e}_{\cdot\cdot})^2}{J-1}}
\end{equation}

where all symbols are as defined earlier.

\subsubsection{Co-location error}

The variability in the true XCO$_2$ contributes to the overall error decomposition in Eq.~\eqref{eq:error_decomposition} via $x_{ijk} - x_{jk}$. While the true field is not available to diagnose this contribution, model fields provide some realism in the spatial variability of XCO$_2$ within coincidence regions. Therefore, observation-based error assessment is augmented by examining model profiles within a day/station coincidence.

We used CarbonTracker (CT2022; see \url{https://gml.noaa.gov/ccgg/carbontracker/}) model-derived vertical profiles of CO$_2$ in the atmosphere extracted at the TCCON site vis-à-vis model profiles extracted at the XCO$_2$ retrieval sites to compute co-location error. This error essentially approximates errors in XCO$_2$ due to a mismatch between latitude, longitude, and time of TCCON measurements and XCO$_2$ observations.

\begin{figure}[h!]
\centering
\includegraphics[width=0.8\textwidth]{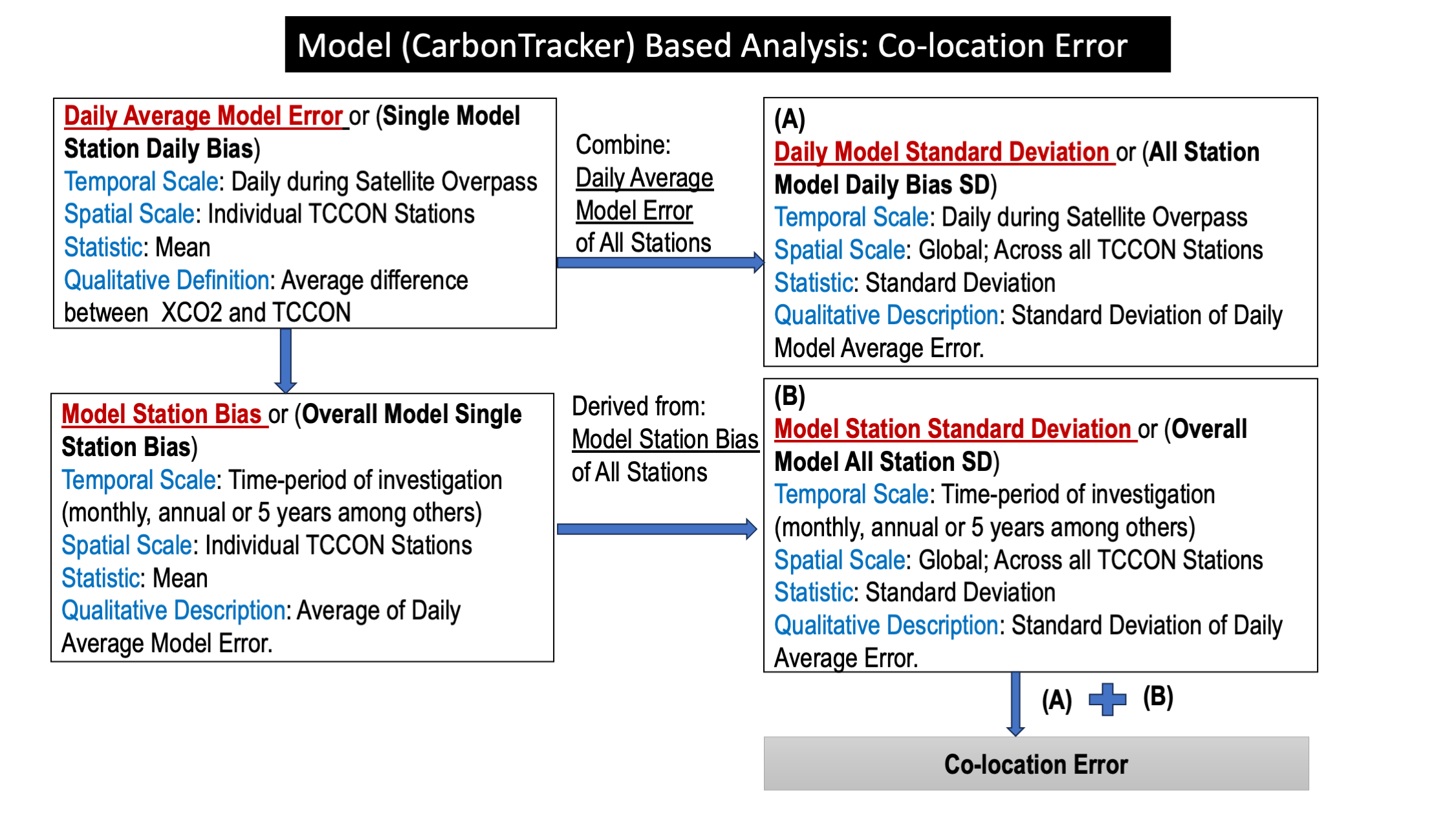}
\caption{A schematic of Carbon Tracker Model-Based Analysis for obtaining Co-location Error. Four intermediary measures for obtaining co-location error are shown in four boxes, and their names are highlighted in red. Note that, as described in the text, only Daily Model Standard Deviation and Model Station Standard Deviation are used for computing co-location errors. The approach to obtain these quantities is same as those shown in Figures~\ref{fig:bias_schematic} and \ref{fig:std_schematic} except for computing Co-location error we use model profiles.}
\label{fig:colocation_schematic}
\end{figure}

For this comparison, a reference profile based on CT2022 at the TCCON site is obtained, and the same is compared with CT2022-based profiles at XCO$_2$ retrieval locations. Two error statistics, i.e., model daily standard deviation and model station bias, are combined to obtain a single measure of co-location error of which the former is representative of variability in transient features of atmospheric CO$_2$ and latter encapsulates the spatial variability of persistent regional co-location differences (e.g. land-sea contrasts for coastal sites). A diagram of sequential steps adopted to obtain co-location error is shown in Figure~\ref{fig:colocation_schematic}.

\subsubsection{Model estimates of TCCON measurements and XCO$_2$ observation}

For obtaining co-location error we compute CT2022-based model profiles i.e., $\mathbf{x}_{mod,ijk}$ corresponding to each $\hat{x}_{ijk}$. The averaging kernel adjustment is applied to the $\mathbf{x}_{mod,ijk}$ using Eq.~\eqref{eq:model_estimate}, which results in CT2022 based estimates of XCO$_2$ i.e., $x_{mc,ijk}$ at spatiotemporal location of $\hat{x}_{ijk}$. Symbolically, this can be given as:
\begin{equation}
\label{eq:model_estimate}
x_{mc,ijk} = \mathbf{h}_{ijk}^T\mathbf{x}_{a,ijk} + \mathbf{h}_{ijk}^T\mathbf{A}_{ijk}(\mathbf{x}_{mod,ijk} - \mathbf{x}_{a,ijk})
\end{equation}

where all quantities are as defined earlier.

To reiterate, for computing co-location error the TCCON profile is not used as a reference. Rather, the model profile i.e., $\mathbf{x}_{mod,jk}$ at the TCCON site is used. This model profile is convolved in a similar way with the matched satellite products, to a reference column, $x_{ref,ijk}$, such that,
\begin{equation}
\label{eq:reference_column}
x_{ref,ijk} = \mathbf{h}_{ijk}^T\mathbf{x}_{a,ijk} + \mathbf{h}_{ijk}^T\mathbf{A}_{ijk}(\mathbf{x}_{mod,jk} - \mathbf{x}_{a,ijk})
\end{equation}

where all quantities are as defined earlier.

Once $x_{mc,ijk}$ and $x_{ref,ijk}$ are computed, we utilize them to compute error statistics. These error statistics are named as daily average model error, model station bias, model station standard deviation, and model daily standard deviation (see Figure~\ref{fig:colocation_schematic}).

\subsubsection{Model error summary statistics and co-location error}

All the model-based error summaries assess the difference between modeled XCO$_2$ at TCCON sites vis-à-vis the XCO$_2$ retrieval locations.

The daily average model error, $\bar{m}_{jk}$, for station $j$ on day $k$ can be computed as:
\begin{equation}
\label{eq:daily_model_error}
\bar{m}_{jk} = \frac{1}{n_{jk}}\sum_{i=1}^{n_{jk}} (x_{mc,ijk} - x_{ref,ijk})
\end{equation}

The model station bias $\bar{m}_{j\cdot}$ for station $j$ is the mean of daily average model errors for that station for a selected time-period and can be given as:
\begin{equation}
\label{eq:model_station_bias}
\bar{m}_{j\cdot} = \frac{1}{N_j} \sum_{k \in \mathbf{d}_j} \bar{m}_{jk}
\end{equation}

Following this, the model station standard deviation $s_{m,b}$, which represents the variability of the individual model station biases for a selected time-period, is computed.
\begin{equation}
\label{eq:model_station_std}
s_{m,b} = \sqrt{\frac{\sum_{j=1}^J (\bar{m}_{j\cdot} - \bar{m}_{\cdot\cdot})^2}{J - 1}}
\end{equation}

Finally, the model daily standard deviation denoted here as $s_{m,d}$, is computed, which represents the variability of the daily average errors, pooled across stations.
\begin{equation}
\label{eq:model_daily_std}
s_{m,d} = \frac{1}{J}\sum_{j=1}^J \sqrt{\frac{1}{N_j - 1} \sum_{k \in \mathbf{d}_j} (\bar{m}_{jk} - \bar{m}_{j\cdot})^2}
\end{equation}

The overall co-location error, denoted here as $s_m$, is the combination of $s_{m,b}$ and $s_{m,d}$.
\begin{equation}
\label{eq:colocation_error}
s_m = \sqrt{s_{m,b}^2 + s_{m,d}^2}
\end{equation}

\subsubsection{Validation Error}

The validation error $s_v = \sqrt{\text{Var}(\epsilon_{val,jk}^*)}$ is an estimate of the error in the TCCON measurements, which has been previously estimated to be 0.4 ppm (1$\sigma$, \citealt{Wunch2011ACP}).

\subsubsection{Systematic error}

Systematic errors are those that remain unchanged when averaging observation errors within a single station/day coincidence. The differences between TCCON and XCO$_2$ retrievals include the co-location errors and errors in TCCON data, i.e., validation error.

From Eq.~\eqref{eq:daily_average_decomposition}, the variance of daily average errors (for large J, $N_j$, and $n_{jk}$) can be estimated as:
\begin{align}
\label{eq:variance_estimation}
\text{Var}(\bar{e}_{jk}) &\approx s_b^2 + s_d^2\\
&\approx s_s^2 + s_m^2 + s_v^2\nonumber
\end{align}

Thus, to compute the satellite-specific systematic error, the co-location and validation error variability are subtracted from the sum of Daily Bias Std ($s_d$), and Station bias Std ($s_b$). Symbolically systematic error can be given as:
\begin{equation}
\label{eq:systematic_error}
s_s = \sqrt{s_b^2 + s_d^2 - s_m^2 - s_v^2}
\end{equation}

where $s_s$ is a systematic error and all other terms are as defined earlier. As noted above, $s_s$ estimates the combined variability of the systematic effects, i.e., $\sqrt{\sigma_\alpha^2 + \sigma_\gamma^2}$.

\subsubsection{Random error}

Random errors are those remaining errors for individual XCO$_2$ observations that do partially decay when multiple observations are averaged. The estimated random error for the retrieved XCO$_2$ observations is constructed from a pooled variance estimate of the differences between satellite and TCCON, after accounting for an analogous estimate from the model profiles.

The variability of random errors ($\epsilon_{ijk}$; Eq.~\eqref{eq:error_decomposition}) is computed from observation level errors of satellite ($e_{ijk}$) and model products ($m_{ijk}$). These observation level errors can be given as:
\begin{align}
\label{eq:observation_errors}
e_{ijk} &= \hat{x}_{ijk} - \bar{x}_{est,jk}\\
m_{ijk} &= x_{mc,ijk} - \bar{x}_{mc,jk}\nonumber
\end{align}

Both these observation level errors have corresponding Std that can be written as:
\begin{align}
\label{eq:observation_std}
s_e &= \frac{1}{J} \sum_{j=1}^J \sqrt{\frac{1}{\sum_{k \in \mathbf{d}_j} n_{jk} - 1} \sum_{k \in \mathbf{d}_j} \sum_{i=1}^{n_{jk}} e_{ijk}^2}\\
s_{m,e} &= \frac{1}{J} \sum_{j=1}^J \sqrt{\frac{1}{\sum_{k \in \mathbf{d}_j} n_{jk} - 1} \sum_{k \in \mathbf{d}_j} \sum_{i=1}^{n_{jk}} m_{ijk}^2}\nonumber
\end{align}

where $s_e$ and $s_{m,e}$ correspond to satellite and model based Std of observation errors, respectively. Once Std of observation errors is computed then, the random error, denoted here as $s_r$, is the square root of the difference in variances,
\begin{equation}
\label{eq:random_error}
s_r = \sqrt{s_e^2 - s_{m,e}^2}
\end{equation}

\subsubsection{Interpretation and Importance of Systematic and Random error}

The random error is useful for small-scale studies, e.g. emissions from power plants, fires, or volcano plumes. When using single observations or aggregating a small number of observations (e.g. the OCO-2 10-sec product), the estimated error $\hat{s}_n$ can be approximated as:
\begin{equation}
\label{eq:estimated_error}
\hat{s}_n = \sqrt{s_s^2 + \frac{s_r^2}{n}}
\end{equation}

where $s_r$ is the random error from Eq.~\eqref{eq:random_error}, $s_s$ is the systematic error from Eq.~\eqref{eq:systematic_error}, and $n$ is the number of observations that are averaged. Eq.~\eqref{eq:estimated_error} can also be used for estimating error when assimilating data into models.

A useful quantity to find is the value of $n$ for which the random error increases the total error by 2\%, or where $\hat{s}_n$/error (infinite measurements) = 1.02.
\begin{equation}
\label{eq:n_threshold}
n = \frac{s_r^2/s_s^2}{1.02^2 - 1}
\end{equation}

For a different fraction, e.g. 1\%, the 1.02 in Eq.~\eqref{eq:n_threshold} would be replaced by 1.01. For MEaSUREs retrieved XCO$_2$ observations over ocean, $s_r^2/s_s^2$ is 0.374; this corresponds to $\sim$ 9.3 observations. In comparison over land, $s_r^2/s_s^2$ is 0.365, which corresponds to $\sim$ 9 observations. For this reason, averaging of at least 10 observations was used for the ``daily average'' cutoff in the next section (see Tables~\ref{tab:land_errors} and \ref{tab:ocean_errors}).

\subsubsection{Prior error}

For comparison, we also evaluate the error of the prior OCO-2 XCO$_2$, which is $\mathbf{h}_{jk}^T\mathbf{x}_{a,ijk}$, following the notation of this section. The OCO-2 prior XCO$_2$ encapsulates some of the large-scale global variability and temporal trend in atmospheric CO$_2$. The prior varies as a function of latitude and time \citep{Crisp2020}, following the approach used for the TCCON network's GGG2020 algorithm \citep{laughner2024}. The purpose of this evaluation is to diagnose the value added by the OCO-2 satellite data and retrieval. For example, if the a priori systematic error is 1.2 ppm and the OCO-2 systematic error is 1.0 ppm, then OCO-2 is adding value over the prior.

The assessment of the prior error is carried out in an analogous fashion to the procedures and Eq.~\eqref{eq:station_bias} to \eqref{eq:estimated_error} defined earlier in this section. In this assessment, the satellite estimate $\hat{x}_{ijk}$ is replaced by the prior XCO$_2$, $\mathbf{h}_{jk}^T\mathbf{x}_{a,ijk}$. Because the prior varies smoothly in space and time, the random error component of the prior error assessment should be nearly zero since the prior XCO$_2$ is nearly constant within a single TCCON coincidence. However, the systematic error of the prior could be large.

\subsection{Comparison of Spatiotemporal Trends in data products}

The XCO$_2$ data products can only be compared with TCCON measurements at specific point locations using a co-location criterion around TCCON sites. While this allows for direct validation, we complemented our statistical error model-based comparison with temporal trend analysis, comparing temporal trends in XCO$_2$ at TCCON sites with trends in XCO$_2$ in the data products.

For broader spatiotemporal analysis, we compared OCO-2 10-sec and MEaSUREs OCO-2 products. Since both are derived from the OCO-2 LtXCO$_2$ native data product, comparing them reveals differences or similarities in their spatiotemporal trends. We exclude comparisons with OCO-2 LtXCO$_2$ as it is the source product for both, making such comparisons redundant. Similarly, we omit comparisons with MEaSUREs OCO-2 and GOSAT because neither OCO-2 LtXCO$_2$ nor OCO-2 10-sec data products incorporate GOSAT data.

To check the similarity between the spatiotemporal distribution of XCO$_2$ concentrations in these products, we spatially discretized observations over 23 TransCom (Atmospheric Tracer Transport Model Intercomparison Project) regions \citep{gurney2002}. This analysis was carried out quarterly from October 1st, 2014, to March 31st, 2023, covering a total of 34 quarters. We chose this discretization approach as it is commonly adopted in data assimilation studies focused on estimating CO$_2$ flux from OCO-2 data \citep{crowell2019}.

\begin{figure}[h!]
\centering
\includegraphics[width=\textwidth]{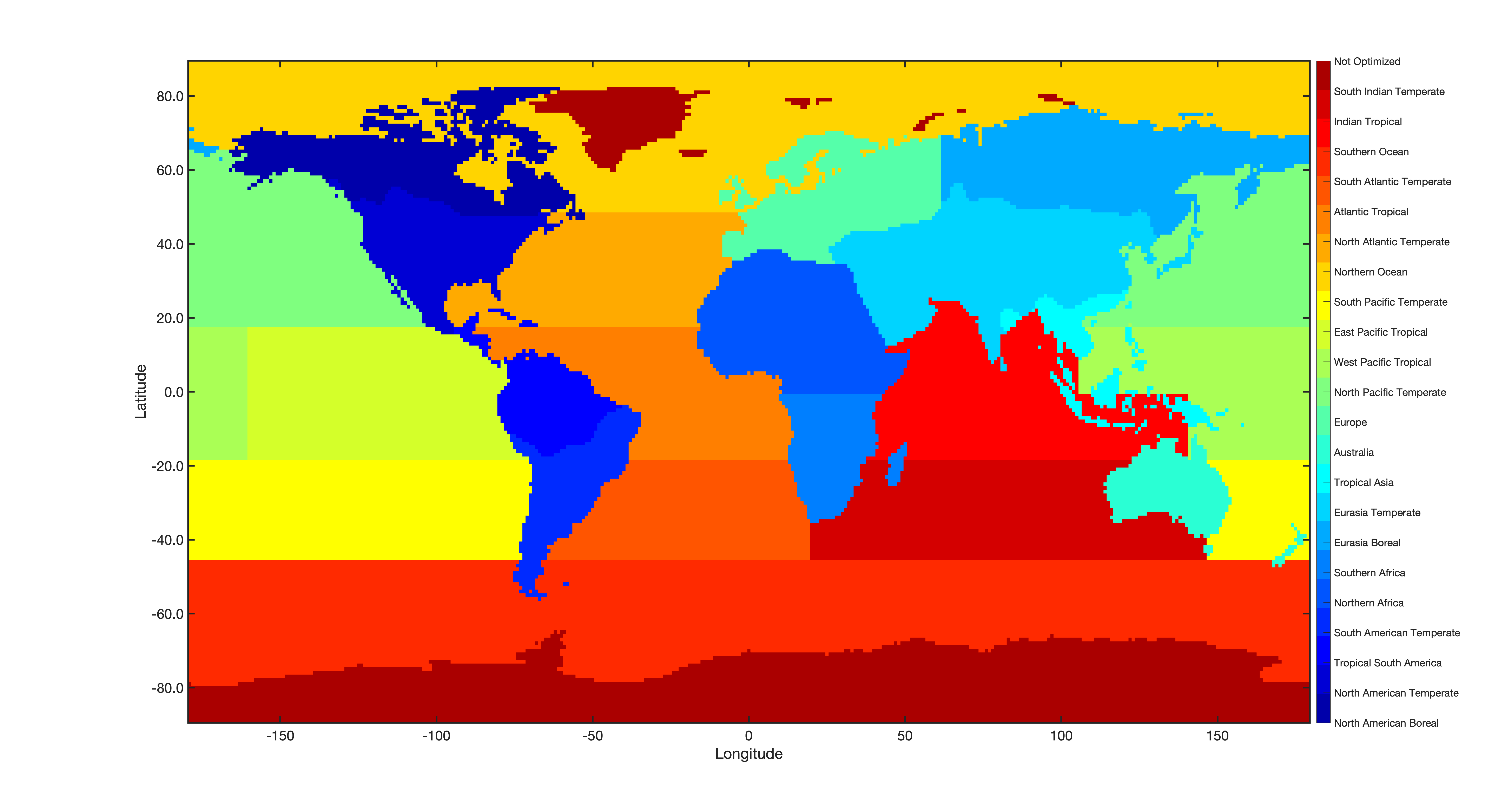}
\caption{Map of 23 TransCom regions used for spatiotemporal analysis used in this work.}
\label{fig:transcom_map}
\end{figure}

\begin{figure}[h!]
\centering
\includegraphics[width=\textwidth]{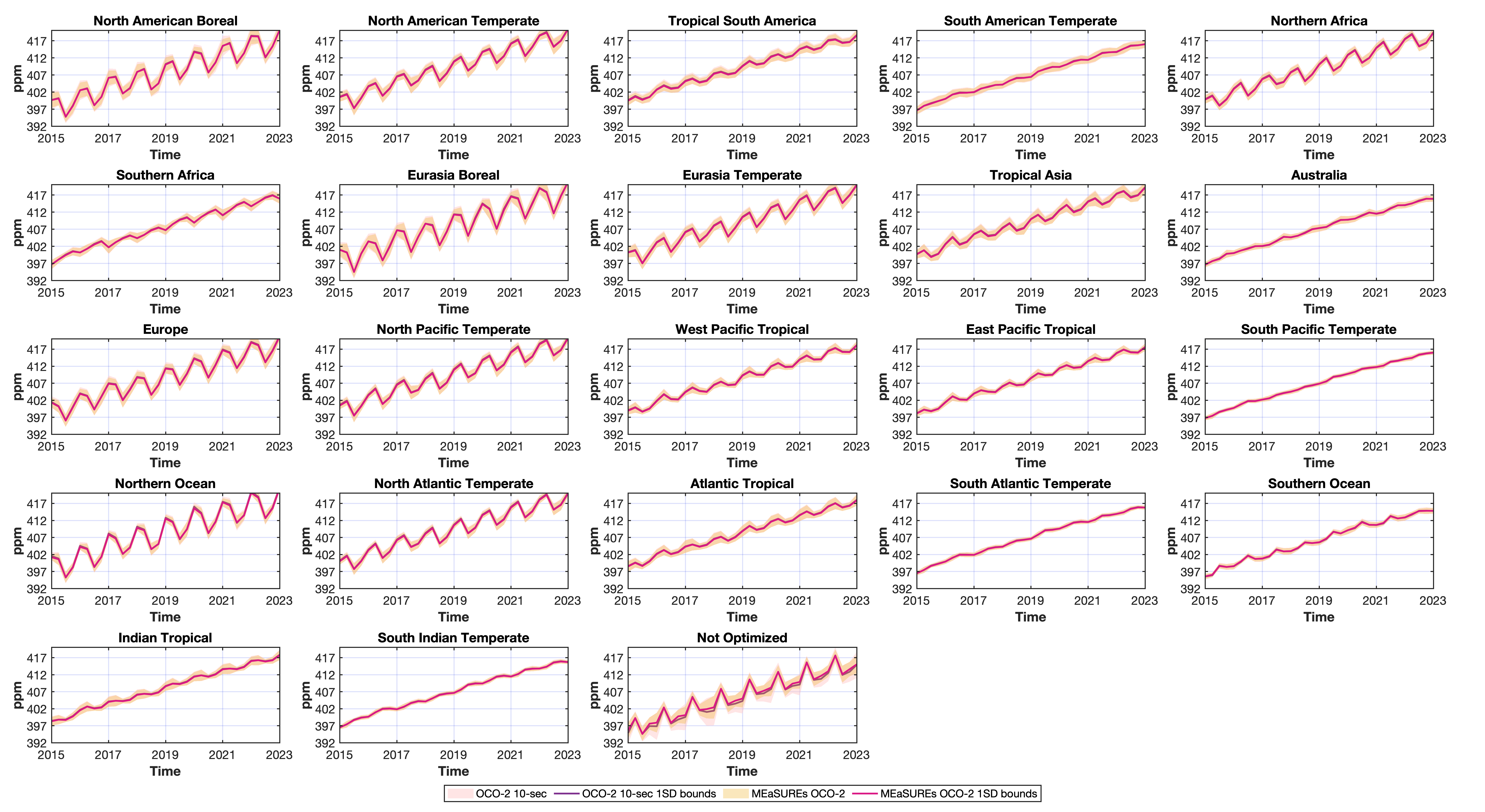}
\caption{Mean and 1-standard deviation (bootstrap standard error) for 23 TransCom regions for 34 quarterly periods.}
\label{fig:transcom_timeseries}
\end{figure}

The products had different numbers of XCO$_2$ observations across all quarters for all TransCom regions as shown in Figure~\ref{fig:transcom_map}. Therefore, we utilized Bootstrap to estimate the mean and standard deviation (SD) of all TransCom regions for 34 quarters, with the time series of this mean and 1 SD bounds shown in Figure~\ref{fig:transcom_timeseries}. The slope and intercept of the linear trend of the time series, with uncertainty for all TransCom regions utilizing the two methods, is reported in Table~\ref{tab:transcom_trends}. Note that while TCCON-based comparison provides in-situ validation against a reference dataset, the inter-comparison of products by TransCom regions demonstrates broader spatiotemporal similarity between the data products.

\section{Results and Discussion}
\label{sec:results}

\subsection{Comparison of MEaSUREs XCO$_2$ products with Total Carbon Column Observing Network (TCCON) data}

Twenty-one TCCON stations were used to compare MEaSUREs and OCO-2 XCO$_2$ products. Table~\ref{tab:tccon_stations} shows these stations' latitude, longitude, and analysis periods.

Systematic error, integrating aspects of model-based co-location error and observation-based error analysis, is the foundation for assessing the compatibility of MEaSUREs XCO$_2$ products with TCCON data. For clarity, the main findings of this comparison are presented in the paper's results tables. At the same time, other analyses concerning TCCON data are discussed in the main text, with corresponding results reported in the supplementary material.

\subsubsection{Systematic Errors over Land and Ocean}

The partitioning of error variability described in Section~\ref{sec:methodology} was carried out for the MEaSUREs products, along with other aggregated and single-sounding OCO-2 products. The results for land products are summarized in Table~\ref{tab:land_errors}, and results for ocean products are summarized in Table~\ref{tab:ocean_errors}. Each row represents a specific data product, along with the timespan of assessment and matchup requirements with the TCCON data. The products included are as follows:

(1) The ``MEaSUREs OCO-2 and GOSAT'' is a 1-degree gridded product that fuses OCO-2 with GOSAT and is compared with TCCON data between January 2015 and March 2020.

(2) The ``MEaSUREs OCO-2'' is a 1-degree gridded product that relies only on OCO-2 observations. It is compared with TCCON data between January 2015 and December 2022.

(3) The ``MEaSUREs OCO-2-fused 2020'' is the MEaSUREs OCO-2 product that is compared with TCCON data for a period between January 2015 and March 2020.

(4) The ``OCO-2'' product is a native OCO-2 LtXCO$_2$ product. It is compared with TCCON data between January 2015 and December 2022, with the requirement that at least ten observations overlap with TCCON sites spatiotemporally during satellite overpass.

(5) The ``OCO-2 10-sec'' product is derived from the OCO-2 LtXCO$_2$ product and compared with TCCON data for the period between January 2015 and December 2022. This product is created by averaging at least ten high-quality observations (based on availability) within a 10-second averaging span.

(6) ``The OCO-2 10-sec*'' product is derived from the native OCO-2 LtXCO$_2$ product. It is compared with TCCON data collected between January 2015 and December 2022.

(7) MEaSUREs OCO-2 prior, the prior error for OCO-2 (before the retrieval). It is compared with TCCON between January 2015 and December 2022.

The systematic errors for land observations (Table~\ref{tab:land_errors}) are 0.96 ppm for both MEaSUREs products. It was lower than 0.99 ppm for the original OCO-2 LtXCO$_2$, OCO-2 10-sec, or MEaSUREs products and 0.98 ppm for the OCO-2 10-sec average product on days with at least ten observations coincidental with TCCON data. This is because sparse OCO-2 LtXCO$_2$ observations, containing less than ten out of 240 possible observations (240 averaged observations is the maximum for 10-sec averages, given OCO-2's 3 Hz cross-scan frequency and eight spatial footprints per cross-scan), generally have higher systematic errors.

Ocean OCO-2 LtXCO$_2$ observations have a lower systematic error (Table~\ref{tab:ocean_errors}) than Land observations across the collection of data products. Thus, OCO-2 LtXCO$_2$ data has a systematic error of 0.62 ppm, whereas this error was 0.68 ppm in the OCO-2 10-sec product. Comparatively, the systematic errors in MEaSUREs products ranged from 0.66 – 0.67 ppm. Similarly, when sparse 10-sec averaged data are included (less than 10 observations per average), the overall systematic error rises to 0.73 ppm for the OCO-2 10-sec product, primarily due to limited sample size after filtering for cloud cover and aerosols. In supplementary material, error summaries at other spatiotemporal granularities (i.e., by year, station, latitude, and season).

The various aggregated products (MEaSUREs, OCO-2 10-sec) exhibit similar magnitudes of systematic and random error for land and ocean products, and random error for these aggregated products is reduced relative to the single-sounding products. These outcomes have implications for the appropriate weighting of satellite XCO$_2$ products in global flux inversions \citep{peiro2022}. The MEaSUREs products provide additional information on spatial correlation across the observations on the same day that can be incorporated into inversions and other downstream analyses.

The decomposition of retrieval error variability summarized here shares similar structure to validation studies for previous OCO-2 data versions \citep{kulawik2019a} and for other satellite greenhouse gas products \citep{kulawik2016}. These studies assembled overpasses with varying sample sizes to estimate a correlated systematic error in combination with the single-sounding random error. This approach differs slightly from the current study, which uses the daily standard deviation in constructing systematic error. The resulting systematic error estimates in the current study are slightly larger over land and ocean compared to the previous approach \citep{kulawik2019b}. At the same time, these systematic error estimates are similar in magnitude to the reported error standard deviation for averaged OCO-2 products in the most recent algorithm versions \citep{jacobs2024, taylor2023}.

\subsubsection{Comparison of systematic error with prior XCO$_2$}

To compare the systematic errors to the size of the signal for the ocean and land observations, we evaluated the OCO-2 prior systematic error versus TCCON using Eq.~\eqref{eq:systematic_error}, the same equation as used to evaluate the OCO-2 LtXCO$_2$, OCO-2 10-sec, or MEaSUREs products systematic error. The prior estimates are shown in the last row of Tables~\ref{tab:land_errors} and \ref{tab:ocean_errors}. The OCO-2 prior XCO$_2$ varies in time and space, particularly with latitude, capturing long-term changes in atmospheric CO$_2$ as well as climatological seasonal cycles \citep{laughner2024}. Thus, real signals and perturbations at regional, intraseasonal scales are not reflected in the prior, and larger systematic errors for the prior would be expected.

We estimated a prior error of 1.84 ppm for land and 0.67 ppm for ocean. Comparing the systematic error for land versus the prior, we see that the prior systematic error (1.84 ppm) is greater than the systematic error for the various products, many of which are just under 1.0 ppm. This result indicates that the various products are capturing local signals effectively even in the presence of these systematic errors. The prior error for the ocean (0.67 ppm) is comparable to the systematic error present in the four products (0.67 ppm). Thus, while the ocean products are more precise than the land products, the precision is on par with the more subtle signals of the atmospheric CO$_2$ perturbations over the oceans. The OCO-2 products have shown expected large-scale perturbations over ocean basins from El Niño events during the last decade \citep{chatterjee2017}. The ocean product assessment and comparisons presented here are also somewhat limited by the lower density of TCCON sites that have coincident ocean satellite products.

\subsubsection{Comparison of seasonal cycle and trend of XCO$_2$ products at TCCON sites}

\begin{figure}[h!]
\centering
\includegraphics[width=\textwidth]{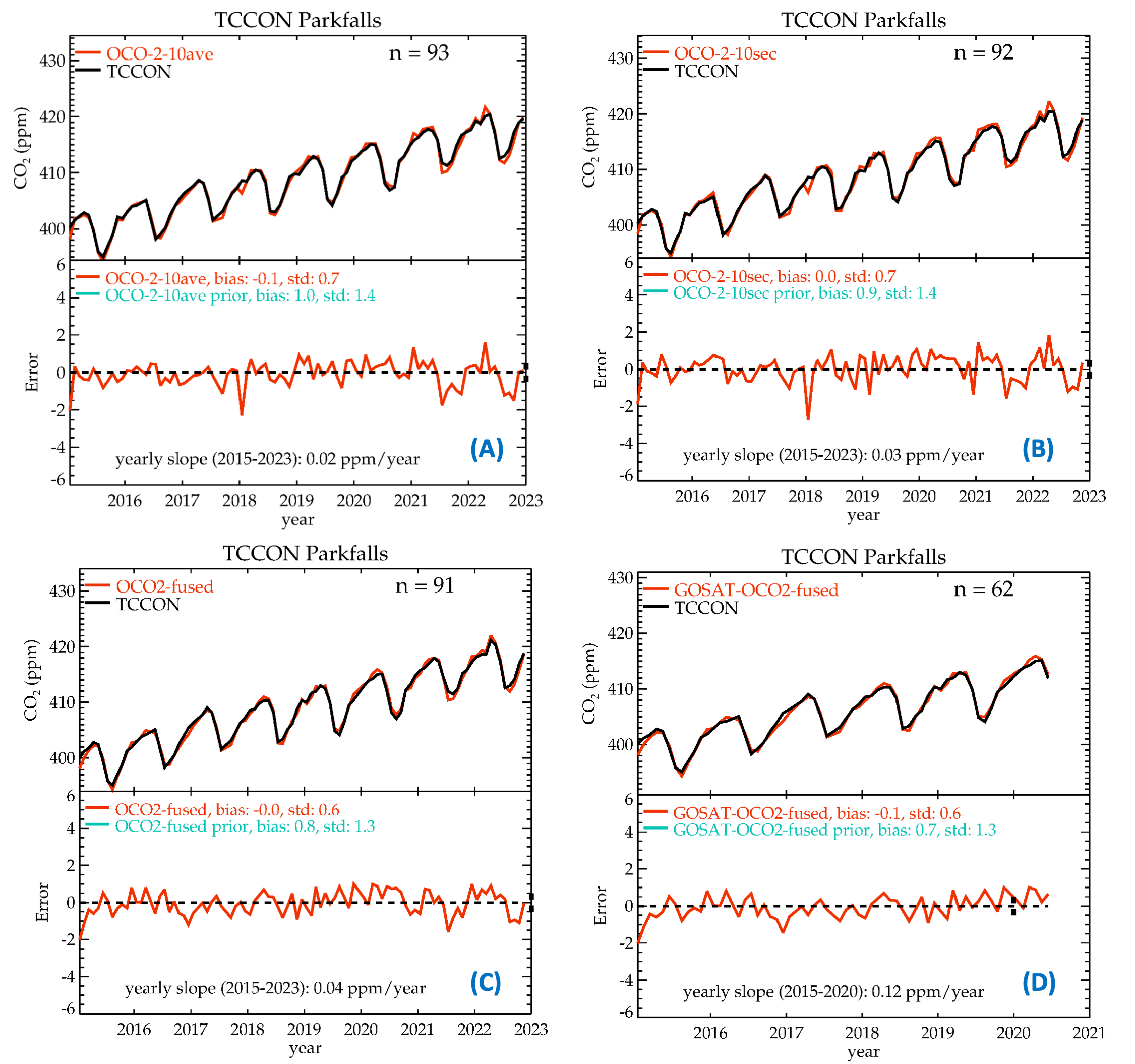}
\caption{The figure compares four XCO$_2$ products at monthly temporal resolution to Park Falls TCCON data within 3°×5° (latitude × longitude) grid cell (i.e., spatial coincidental criteria). Four panels show a comparison of TCCON data with: (A) OCO-2 LtXCO$_2$, (B) OCO-2 10-sec, (C) MEaSUREs OCO-2, and (D) MEaSUREs OCO-2 and GOSAT.}
\label{fig:parkfalls_comparison}
\end{figure}

\begin{figure}[h!]
\centering
\includegraphics[width=\textwidth]{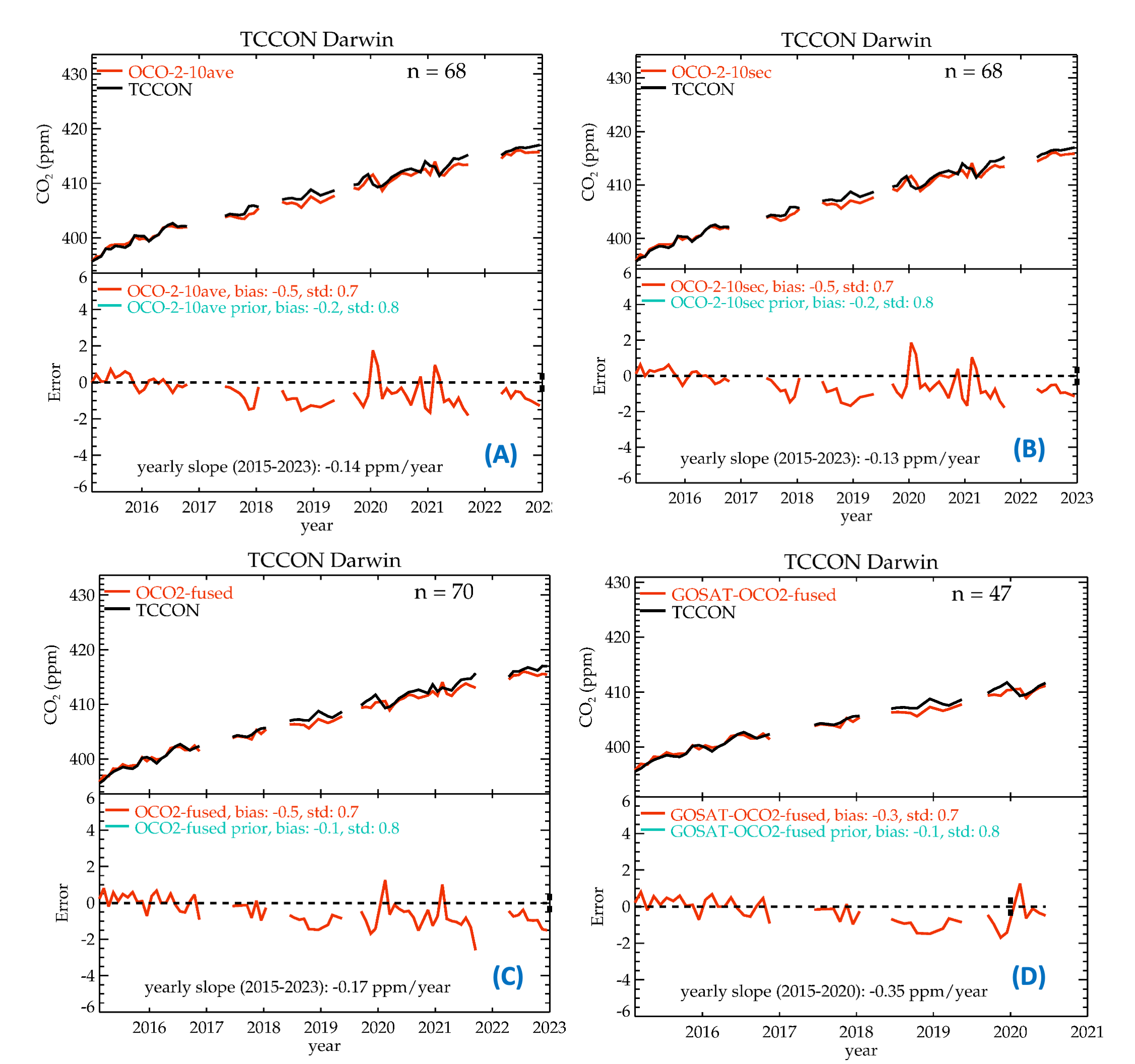}
\caption{Monthly comparison of XCO$_2$ products to Darwin TCCON data within 3°×5° (latitude × longitude) grid cell (i.e., spatial coincidental criteria). Four panels show comparison (A) OCO-2 LtXCO$_2$, (B) OCO-2 10-sec, (C) MEaSUREs OCO-2, and (D) MEaSUREs OCO-2 and GOSAT.}
\label{fig:darwin_comparison}
\end{figure}

The XCO$_2$ data in all MEaSUREs products and OCO-2 and OCO-2 10-sec product were very close to TCCON data. As an example, at monthly resolution, the trend in both TCCON data for Park Falls (Figure~\ref{fig:parkfalls_comparison}) and Darwin site (Figure~\ref{fig:darwin_comparison}), which are sites that are located at an interface of Land and Ocean, had similar bias (difference between the mean temporal trend of TCCON observations and TCCON data) and variability as the observations. This confirms that all products have the same information as those in TCCON data. At monthly resolution, bias in all the products was nearly zero, indicating that any conclusion about the temporal trend of XCO$_2$ concentration, such as its growth rate, gleaned from these products would be no different from those obtained from TCCON data. Note that MEaSUREs products are gridded products. Therefore, the number of XCO$_2$ observations in these products was lower than in the other two. However, even with fewer observations, MEaSUREs products contain the same information as those in the other two products.

\subsection{Intercomparison between data products}

Mean XCO$_2$ concentrations between the MEaSUREs OCO-2 and OCO-2 10-sec data products showed no significant differences across all TransCom regions over 34 quarterly periods, as confirmed by bootstrap uncertainty bounds shown in Figure~\ref{fig:transcom_timeseries}.

The linear trend slopes and uncertainties for these means (Table~\ref{tab:transcom_trends}) further support this finding. The TCCON-based trend analysis results demonstrate that both data products consistently capture carbon dioxide trends at point locations and regional scales.

\section{Conclusion}
\label{sec:conclusion}

A primary goal of this work was to develop a hierarchical statistical framework for quantifying systematic and random errors when comparing XCO$_2$ measurements against reference data. Our analysis demonstrates that the gridded MEaSUREs dataset maintains signal-to-noise characteristics comparable to other products, including the native OCO-2 LtXCO$_2$ data, as verified through TCCON and TransCom-based analysis. These findings support the use of MEaSUREs products in scientific applications without concerns about XCO$_2$ measurement quality. While demonstrated here for XCO$_2$ comparisons with TCCON data, this hierarchical framework is versatile and applicable to evaluating any satellite-based product against reference datasets.

The MEaSUREs XCO$_2$ products are distinctive in providing covariance and precision matrices that characterize XCO$_2$ distribution based on spatiotemporal correlation. To our knowledge, these are the first satellite products to include full error covariances with off-diagonal terms. This feature enables data assimilation systems to account for off-diagonal measurement errors when estimating fluxes from XCO$_2$ measurements.

Future improvements to the error framework could incorporate varying overall bias and correlation in station-specific bias, daily overpass, and retrieval errors (see Eq.~\eqref{eq:error_decomposition} and \eqref{eq:daily_average_decomposition}). Additionally, MEaSUREs products could be enhanced by incorporating temporal correlation alongside existing spatial correlation, though this would increase computational complexity. These potential enhancements warrant further investigation in future research.

\section*{Acknowledgements}

A portion of this research has been carried out at the Jet Propulsion Laboratory, California Institute of
Technology, under a contract with NASA (contract no. 80NM0018F0527) as part of the 2017 Making Earth System Data Records for Use in Research Environments (MEaSUREs) program. TCCON sites at Tsukuba, Rikubetsu, and Burgos are supported in part by the GOSAT series project. Burgos is supported in part by the Energy Development Corporation Philippines. The authors declare no potential conflicts of interest with respect to the research, authorship, and/or publication of this article.

\section*{Data Availability}

All the data used in this work is available online. The data from Total Carbon Column Observing Network (TCCON) is \url{https://tccondata.org/}. TCCON data can freely uses for research, however, users are required to follow TCCON data use guidelines as described on webpage \url{https://tccon-wiki.caltech.edu/Main/DataUseGuidelines}. This webpage also includes a link to the data license information that is described on webpage \url{https://tccon-wiki.caltech.edu/Main/DataLicense}. We only provide a general reference to the TCCON as mentioned on TCCON website. References for data for individual sites can be generated from \url{https://tccondata.org/metadata/siteinfo/genbib/}.

OCO-2 LtXCO$_2$ v11.1 is freely available from the Goddard Earth Sciences
Data and Information Services Center 
(GES DISC; \href{https://disc.gsfc.nasa.gov/datasets/OCO2_L2_Lite_FP_11.1r/summary?keywords=OCO-2}{dataset page}).
The two gridded products are available from 
\href{https://disc.gsfc.nasa.gov/datasets/OCO2GriddedXCO2_3/summary?keywords=OCO-2\%20XCO2}{OCO-2 Gridded XCO$_2$ v3} 
and 
\href{https://disc.gsfc.nasa.gov/datasets/MultiInstrumentFusedXCO2_3/summary?keywords=MultiInstrumentFusedXCO2_3}{Multi-Instrument Fused XCO$_2$ v3}.
The OCO-2 10-sec product is hosted on 
\href{https://zenodo.org/records/14866520}{Zenodo} 
and is also available from 
\href{https://gml.noaa.gov/aftp/user/andy/OCO-2/}{NOAA GML}.

% Figures

% Tables
\begin{table}[h!]
\centering
\caption{TCCON stations for validation of MEaSUREs, OCO-2, and OCO-2 10-sec products. Note that columns Begin Year and End Year indicate the period of comparison. The stations identified as both land and ocean sites are located on coastlines or islands; therefore, they are considered both land and ocean sites. *Water data from the Park Falls site is not available for MEaSUREs products as 1° MEaSUREs grid did not cover this region.}
\label{tab:tccon_stations}
\begin{tabular}{clcccccc}
\toprule
No & Station & Latitude & Longitude & Begin Year & End Year & Land & Ocean \\
\midrule
1 & Eureka & 80 & -86.4 & 2015 & 2020 & X &  \\
2 & NyAlesund & 78.9 & 11.9 & 2015 & 2022 & X & X \\
3 & Sodankyla & 67.4 & 26.6 & 2015 & 2022 & X &  \\
4 & Easttroutlake & 54.4 & -105 & 2016 & 2022 & X &  \\
5 & Bremen & 53.1 & 8.8 & 2015 & 2021 & X &  \\
6 & Karlsruhe & 49.1 & 8.4 & 2015 & 2022 & X &  \\
7 & Orleans & 48 & 2.1 & 2015 & 2022 & X & X \\
8 & Garmisch & 47.5 & 11.1 & 2015 & 2022 & X &  \\
9 & Parkfalls & 45.9 & -90.3 & 2015 & 2022 & X & X* \\
10 & Rikubetsu & 43.5 & 143.8 & 2015 & 2021 & X & X \\
11 & Lamont & 36.6 & -97.5 & 2015 & 2022 & X &  \\
12 & Tsukuba125 & 36 & 140.1 & 2015 & 2021 &  & X \\
13 & Nicosia & 35.1 & 33.4 & 2019 & 2021 & X & X \\
14 & Dryden & 35 & -117.9 & 2015 & 2022 & X &  \\
15 & Saga & 33.2 & 130.3 & 2015 & 2022 & X & X \\
16 & Hefei & 31.9 & 119.2 & 2015 & 2020 & X &  \\
17 & Burgos & 18.5 & 120.7 & 2017 & 2021 & X & X \\
18 & Manaus & -3.2 & -60.6 & 2015 & 2015 & X &  \\
19 & Darwin & -12.4 & 130.9 & 2015 & 2022 & X & X \\
20 & Wollongong & -34.4 & 150.9 & 2015 & 2022 & X & X \\
21 & Lauder125 & -45 & 169.7 & 2015 & 2022 & X & X \\
\bottomrule
\end{tabular}
\end{table}

\begin{table}[h!]
\centering
\caption{Error statistics of different XCO$_2$ products over Land derived from comparison with 20 TCCON sites. Note that MEaSUREs products generally have fewer observations than OCO-2 and OCO-2 10-sec products as they are gridded spatially at 1°×1° resolution.}
\label{tab:land_errors}
\small
\begin{tabular}{lcccccccc}
\toprule
LAND & Daily & Overall & Bias & Daily & Co- & Validation & System- & Random \\
 & Obs. & Bias & Stdev & Stdev & location & Error & atic Error & Error \\
 &  & (ppm) & (ppm) & (ppm) & Error & (ppm) & (ppm) & (ppm) \\
 &  &  & or Station &  & (ppm) &  &  &  \\
 &  &  & Bias &  &  &  &  &  \\
\midrule
MEaSUREs & 2377 & -0.19 & 0.4 & 1.03 & 0.37 & 0.4 & 0.96 & 0.58 \\
OCO-2 and &  &  &  &  &  &  &  &  \\
GOSAT &  &  &  &  &  &  &  &  \\
MEaSUREs & 3413 & -0.2 & 0.42 & 1.03 & 0.39 & 0.4 & 0.96 & 0.58 \\
OCO-2 &  &  &  &  &  &  &  &  \\
MEaSUREs & 2246 & -0.18 & 0.36 & 1.04 & 0.36 & 0.4 & 0.96 & 0.54 \\
OCO2-fused- &  &  &  &  &  &  &  &  \\
2020 &  &  &  &  &  &  &  &  \\
OCO-2 & 4717 & -0.12 & 0.53 & 1.01 & 0.39 & 0.4 & 0.99 & 0.98 \\
LtXCO2 &  &  &  &  &  &  &  &  \\
OCO-2 & 4271 & -0.29 & 0.49 & 0.99 & 0.29 & 0.4 & 0.98 & 0.52 \\
10-sec &  &  &  &  &  &  &  &  \\
OCO-2 10- & 4778 & -0.31 & 0.44 & 1.16 & 0.29 & 0.4 & 1.13 & 0.73 \\
sec* &  &  &  &  &  &  &  &  \\
MEaSUREs & 3413 & 0.00 & -0.74 & 1.42 & 1.45 & 0.39 & 0.4 & 1.84 \\
OCO-2 Prior &  &  &  &  &  &  &  &  \\
\bottomrule
\end{tabular}
\end{table}

\begin{table}[h!]
\centering
\caption{Error statistics of different XCO$_2$ products over Ocean derived from comparison with 12 TCCON sites. Out of these 12 sites, only 11 were available for comparison for MEaSUREs products as MEaSUREs data does not have water observations near Park Falls (over the Great Lakes), which are found in the other products.}
\label{tab:ocean_errors}
\small
\begin{tabular}{lcccccccc}
\toprule
Ocean & Daily & Overall & Bias & Daily & Co- & Validation & System- & Random \\
 & Obs. & Bias & Stdev & Stdev & location & Error & atic Error & Error \\
 &  & (ppm) & (ppm) & (ppm) & Error & (ppm) & (ppm) & (ppm) \\
 &  &  & or Station &  & (ppm) &  &  &  \\
 &  &  & Bias &  &  &  &  &  \\
\midrule
MEaSUREs & 718 & -0.1 & 0.34 & 0.78 & 0.33 & 0.4 & 0.67 & 0.41 \\
OCO-2 and &  &  &  &  &  &  &  &  \\
GOSAT &  &  &  &  &  &  &  &  \\
MEaSUREs & 1031 & -0.12 & 0.35 & 0.77 & 0.32 & 0.4 & 0.67 & 0.38 \\
OCO-2 &  &  &  &  &  &  &  &  \\
MEaSUREs & 686 & -0.1 & 0.34 & 0.78 & 0.34 & 0.4 & 0.66 & 0.37 \\
OCO2-fused- &  &  &  &  &  &  &  &  \\
2020* &  &  &  &  &  &  &  &  \\
OCO-2 & 1435 & -0.16 & 0.35 & 0.76 & 0.37 & 0.4 & 0.62 & 0.51 \\
OCO-2 & 1235 & -0.19 & 0.34 & 0.77 & 0.28 & 0.4 & 0.68 & 0.33 \\
10-Second &  &  &  &  &  &  &  &  \\
OCO-2 10- & 1391 & -0.19 & 0.37 & 0.8 & 0.28 & 0.4 & 0.73 & 0.4 \\
Second** &  &  &  &  &  &  &  &  \\
MEaSUREs & 1031 & 0.32 & -0.12 & 0.85 & 0.77 & 0.32 & 0.4 & 0.67 \\
OCO-2 Prior &  &  &  &  &  &  &  &  \\
\bottomrule
\end{tabular}
\end{table}

\clearpage
\begin{longtable}{lcccc}
\caption{Intercept and slope of 10-second and MEaSUREs product.} \\
\label{tab:transcom_trends} \\
\toprule
TransCom & Intercept & Slope & Intercept & Slope \\
Regions & 10-second & 10-second & MEaSUREs & MEaSUREs \\
 &  &  & product & product \\
\midrule
\endfirsthead
\toprule
TransCom & Intercept & Slope & Intercept & Slope \\
Regions & 10-second & 10-second & MEaSUREs & MEaSUREs \\
 &  &  & product & product \\
\midrule
\endhead
North American Boreal & 397.33 ± 1.02 & 0.62 ± 0.05 & 397.28 ± 1.00 & 0.62 ± 0.05 \\
North American Temperate & 399.14 ± 0.73 & 0.61 ± 0.04 & 398.99 ± 0.72 & 0.61 ± 0.04 \\
Tropical South America & 399.30 ± 0.29 & 0.58 ± 0.01 & 399.13 ± 0.29 & 0.59 ± 0.01 \\
South American Temperate & 397.03 ± 0.15 & 0.59 ± 0.01 & 396.99 ± 0.13 & 0.59 ± 0.01 \\
Northern Africa & 398.94 ± 0.62 & 0.60 ± 0.03 & 398.81 ± 0.61 & 0.60 ± 0.03 \\
Southern Africa & 397.55 ± 0.30 & 0.59 ± 0.02 & 397.50 ± 0.27 & 0.59 ± 0.01 \\
Eurasia Boreal & 397.82 ± 1.10 & 0.62 ± 0.06 & 397.79 ± 1.08 & 0.62 ± 0.06 \\
Eurasia Temperate & 398.81 ± 0.72 & 0.61 ± 0.04 & 398.68 ± 0.70 & 0.61 ± 0.04 \\
Tropical Asia & 399.12 ± 0.41 & 0.60 ± 0.02 & 398.99 ± 0.42 & 0.60 ± 0.02 \\
Australia & 397.00 ± 0.16 & 0.60 ± 0.01 & 396.97 ± 0.14 & 0.60 ± 0.01 \\
Europe & 398.56 ± 0.92 & 0.61 ± 0.05 & 398.43 ± 0.90 & 0.61 ± 0.05 \\
North Pacific Temperate & 399.33 ± 0.76 & 0.61 ± 0.04 & 399.16 ± 0.74 & 0.61 ± 0.04 \\
West Pacific Tropical & 398.46 ± 0.35 & 0.60 ± 0.02 & 398.35 ± 0.34 & 0.60 ± 0.02 \\
East Pacific Tropical & 398.10 ± 0.26 & 0.60 ± 0.01 & 398.01 ± 0.27 & 0.60 ± 0.01 \\
South Pacific Temperate & 396.87 ± 0.13 & 0.60 ± 0.01 & 396.84 ± 0.12 & 0.60 ± 0.01 \\
Northern Ocean & 398.33 ± 1.18 & 0.62 ± 0.06 & 398.18 ± 1.12 & 0.62 ± 0.06 \\
North Atlantic Temperate & 399.20 ± 0.73 & 0.60 ± 0.04 & 399.04 ± 0.71 & 0.60 ± 0.04 \\
Atlantic Tropical & 398.40 ± 0.28 & 0.59 ± 0.01 & 398.29 ± 0.29 & 0.59 ± 0.01 \\
South Atlantic Temperate & 396.95 ± 0.16 & 0.60 ± 0.01 & 396.91 ± 0.16 & 0.60 ± 0.01 \\
Southern Ocean & 395.91 ± 0.28 & 0.61 ± 0.01 & 395.94 ± 0.25 & 0.61 ± 0.01 \\
Indian Tropical & 398.01 ± 0.20 & 0.60 ± 0.01 & 397.91 ± 0.21 & 0.60 ± 0.01 \\
South Indian Temperate & 396.91 ± 0.18 & 0.60 ± 0.01 & 396.88 ± 0.17 & 0.60 ± 0.01 \\
Not Optimized & 395.25 ± 0.85 & 0.61 ± 0.04 & 395.73 ± 0.76 & 0.61 ± 0.04 \\
\bottomrule
\end{longtable}

% Bibliography
\bibliographystyle{ametsocV6}
\bibliography{references}

\end{document}